\documentstyle[11pt,aasms,psfig]{article}
\begin{document}
\newcommand{\etal}{{\it et~al.}\ }%

\title{MOLECULES AT HIGH REDSHIFT: THE EVOLUTION OF THE COOL PHASE OF
PROTOGALACTIC DISKS}

\author{Colin A. Norman$^{1, 2}$ and Marco Spaans$^{1}$ } 
\affil{Department of Physics and Astronomy,
Johns Hopkins University$^{1}$ and Space Telescope Science Institute$^{2}$ }
\authoraddr{3400 N Charles Street, Baltimore, MD 21218 }


\begin{abstract}
 
We study the formation of molecular hydrogen, after the epoch of
re-ionization, in the context of canonical galaxy formation theory due
to hierarchical clustering. There is an initial epoch of $H_2$
production in the gas phase through the $H^-$ route which ends at a
redshift of order unity. We assume that the fundamental units in the
gas phase of protogalaxies during this epoch are similar to diffuse
clouds found in our own galaxy and we restrict our attention to
protogalactic disks although some of our analysis applies to
multi-phase halo gas. Giant molecular clouds are not formed until
lower redshifts. Star formation in the protogalactic disks can become
self-regulated. The process responsible for the feedback is heating of the gas 
by the internal stellar radiation field which can dominate the background
radiation field at various epochs. If the gas is heated to above
$2000 - 3000 K$ the hydrogen molecules are collisionally dissociated
and we assume that in their absence the star formation process is
strongly suppressed due to insufficient cooling. As we demonstrate by
analysis of phase diagrams, the H$_2$-induced cool phase disappears. A priori,
the cool phase with molecular hydrogen cooling can only achieve
temperatures $ \geq 300 K$. Consequently, it is possible to define a
maximum star formation rate during this epoch. Plausible estimates
give a rate of $ \lesssim 0.2-2 M_{\odot} yr^{-1}$ for condensations
corresponding to 1$\sigma$ and 2$\sigma$ initial density fluctuations. For more
massive structures, this limit is relaxed and in agreement with observations
of high redshift galaxies. Therefore, the
production of metals and dust proceeds slowly in this phase. This
moderate epoch is terminated by a phase transition to a cold dense and
warm neutral/ionized medium once the metals and dust have increased
to a level $Z\approx 0.03-0.1Z_{\odot}$. Then: (1) atoms and molecules
such as C, O and
CO become abundant and cool the gas to below $300 K$ ; (2) the dust abundance
has become sufficiently high to allow shielding of the molecular gas and;
(3) molecular hydrogen formation can occur rapidly on grain
surfaces. This phase transition occurs at a redshift of approximately 1.5,
with a fiducial range of $1.2\le z\le 2$, and
initiates the rapid formation of molecular species, giant molecular clouds and
stars. The delayed initiation of the cold phase in the interstellar
medium of protostellar disks at a metallicity of $Z \lesssim 0.1
Z_{\odot}$ is consequently a plausible physical reason that the
formation phase of the stellar disks of the bulk of the galaxies
only occurs at a redshift of order unity.
The combination of feedback and a phase transition provides a natural
resolution of the G-dwarf problem.\hfill\break
\noindent
{\it subject headings}: cosmology: theory - galaxies: formation -
galaxies: evolution - molecular processes

\end{abstract}

\section{Introduction}

Despite numerous searches for molecules at high redshift, the positive
results are sparse. The only confirmed observations of molecules at
high redshift seen in emission are associated with two gravitational
lenses: FSC~10214~+ 4724 (Scoville \etal 1995, Frayer 1995) and the
Cloverleaf (Barvainis \etal 1995). In absorption, $H_2$ has been seen
in PKS~0528~- 250 (Foltz \etal~1988) at a column density of
$10^{18}$~cm$^{-2}$, but in no other damped Lyman Alpha system. This
is surprising since the column densities of atomic hydrogen inferred
for damped Lyman Alpha systems are $\sim 10^{21}$~cm$^{-2}$.  In
comparison with diffuse clouds in our own galaxy such columns seem
sufficient to produce $H_2$. Galactic diffuse clouds cover the disk
with a covering factor between 20--50\% where the lower limit is from
the Copernicus data (Savage \etal 1977) and the upper limit comes from
assuming all the infrared Cirrus is associated with diffuse clouds.
Other molecular species including HCO$^+$, HCN, CO, C$^{18}$O, HNC,
CN, CS and H$_2$CO have been seen below redshift $\sim 1$ in the
millimeter band (Combes \& Wiklind 1996). These absorption techniques
are effective for any redshift, provided a bright millimeter continuum
background exists. Nevertheless, all the absorbers known to date are
in lensed systems with correspondingly small impact parameters.

At lower redshift $z \leq 0.2$ there are well established observations
of many molecular species in emission in the nuclei of starburst
galaxies (c.f Scoville \& Soifer 1991). For large nearby galaxies
such as M31 and NGC~891 (c.f.~Young 1990) CO maps have
been obtained. For our own galaxy, about a hundred different molecular
species have been detected in interstellar space. Many di- and
tri-atomic molecules can be observed toward diffuse clouds through the
visual and millimeter absorption lines they produce in the continua of
nearby stars. Some quasars ($\sim 30$\%) used as millimeter callibration 
sources are known to exhibit absorption lines as well
(de Geus \etal 1996).

These data pose interesting questions about the prevailing chemical
balance and the physical structure of the ambient medium as a function
of redshift (c.f.\ Norman \& Braun 1996). Previous work on this
subject has focused on very high redshift in the epoch prior to
reionization where light-element hydrides such as $DH$ and $LiH$ have
been discussed in the context of observations of deuterium abundances
and faint structure in the cosmic background radiation field (Maoli
\etal 1996, Dalgarno \etal 1996, Stancil \etal 1996). The importance
of $H_2$ cooling for the formation of the first stars has been
discussed by many authors (c.f.\ Haiman \etal 1996, Tegmark \etal
1996). Although the self-consistent formation of $H_2$ above a
redshift of $z \sim 6$ due to gas phase processes yields very low
abundances of molecular hydrogen $H_2 / H \sim 10^{-6}$ (Lepp \&
Shull 1984), these may be sufficient to sustain the collapse of
initial perturbations. In the post re-ionization epoch, the problem of
$H_2$ formation has been discussed by Black \etal (1987) for the
QSO~PHL~957---a damped Lyman Alpha system with an atomic $HI$ column
density of $2.5\times 10^{21}$~cm $^{-2}$---where the $H_2$ abundance
was more than five orders of magnitude lower than values
representative of our own galaxy (c.f.~Levshakov \& Varshalovich 1985,
Lanzetta \etal 1989). Their analysis indicated that two essential
parameters are: (1)~the background radiation field which is inferred
to be an order of magnitude higher than typical intergalactic values at
$z=2.3$, and (2)~the dust-to-gas ratio which is constrained to be no
more than $\sim 1/3$ of typical galactic values. Many new results have
recently become available on the UV background radiation field (cf.\
Haardt \& Madau 1995), the dust content of damped Lyman Alpha
systems (Pettini \etal 1994, Fall \& Pei 1994) and the merging history of
dark matter halos (Kauffmann \& White 1993). QSO absorption line
studies have also made remarkable progress in the last few years. The
high redshift ($z>2$) dependence of the $HI$ column density
distribution function contains valuable information on the early
stages of structure formation during which large systems are likely to
be in the process of aggregating themselves from smaller
sub-units. The obvious candidates to focus on are those systems with
$HI$ column densities above $N(HI) \geq 10^{21}$~cm$^{-2}$ (Wolfe
1995, Storrie-Lombardi \etal 1994). These systems are often associated
with protostellar disks and are believed to contain a reasonable
fraction of the baryons of the universe.

These new developments motivate us to re-investigate the formation of
molecular hydrogen after re-ionization and to assess its role in the
standard galaxy formation theory due to hierarchical clustering and
the subsequent formation of stars. If the latter process is to occur, a
diffuse cool ($T \lesssim 500 K$) phase needs to be supported which
ultimately leads to the formation of dense molecular clouds and
stars. Obviously, the formation of stars produces an internal
radiation field which will heat its surroundings and influence the
chemical balance. For star formation to continue, the cool phase
should persist. In fact, this process will determine the maximally
sustained star formation rate. If the associated time scale is long
(comparable to a Hubble time), the abundance of metals and dust will
increase slowly until a threshold is reached at which the cooling is
dominated by atoms and ions like O and C$^+$, as well as molecular
species like CO. In addition, H$_2$ formation on grains proceeds more
rapidly and dominates in regions where the bulk of the material is in
neutral form. Also, the increased columns of dust absorb the radiation
of newly formed stars and sustain a star formation cycle as in our own
galaxy. The onset of the bulk of the star formation at low redshifts
and moderate metallicity is consistent with current constraints on
disk formation. In this paper we have restricted our attention to
protogalactic disks since they are the likely sites of the cold
molecular phase we are focusing on. Multi-phase halos have been
discussed extensively for galaxy formation (c.f.\ Kang et al 1991,
Ikeuchi \& Norman 1991) and our results in sections 2.5, 3, 4, and 5 are
relevant for such studies although we do not pursue them further here.

This paper is organized as follows. In Section 2, the formation and
subsequent evolution of structure in a CDM-dominated Universe is
discussed, and the redshift dependence of the UV background radiation
field, the stellar radiation field, the gas pressure, the radiation
pressure and the ionization parameter are presented. The gas-phase
epoch of molecular hydrogen formation is calculated in Section~3
including the production of metals and dust. The feedback effect due
to star formation is analyzed in Section 4 and upper limits to the
star formation rate and metal and dust production rate are
derived. In section 5 we discuss the transition to a multi-phase ISM
and the onset of rapid star formation. The epoch when H$_2$ formation on
grains becomes important is
discussed in Section 6. The formation of metals and subsequent
molecular cooling to massive molecular clouds is also discussed there.
Section 7 discusses the implications for the study of the interstellar
medium in protogalaxies and star formation, as well as the G-dwarf
problem. We associate
the onset of star formation in disks with both a redshift, $z \sim
1-2$, and a metallicity, $Z \sim 0.03-0.1$ as also indicated in a perceptive
paper by Wyse and Gilmore (1988). Where possible we give reasonable
analytical estimates and then use the full (numerical) details of the
recently calculated metagalactic and stellar radiation fields in our
calculations.

\section{Basic Cosmological Model}

\subsection{Galaxy Formation}

For the present investigation the specific details of structure formation are
of minor importance and
we adopt the semi-analytical relations presented by White \& Frenk (1991) 
which agree well with more detailed N-body simulations (cf.\ Lacy
\& Cole 1993, 1994; Kauffmann \etal 1993, Heyl \etal 1995). In the numerical
calculations, we also include the merging history of dark matter halos through
the conditional probability that a halo of mass $M_0$ at redshift $z_0$ has
previously been in a halo of mass $M_1$ at $z_1$, as presented in
Kauffmann \& White (1993).

Using simple top-hat models, the collapse to virial equilibrium of a
perturbation that has become non-linear at redshift $z$ results in a
density enhancement $\delta$. In CDM scenarios the fiducial value is
$\delta =178$ (Narayan \& White 1988). We assume that for
protogalactic disks to form, cooling takes place and a fraction of the
gas originally contained in the mass perturbation collapses to a
centrifugally supported disk with a collapse factor of $\lambda^{-1}$,
where $\lambda$ is the spin parameter from tidal torques (cf.\ Peebles
1993) with a canonical value of $\lambda =0.07$. Note that there are
details of the cooling and infall to the protogalactic disk that we
have omitted but they are not essential to the argument given
here. For a density perturbation of mass $M$, the resulting column
density is given by
\begin{equation}
N = \left({4 \pi \over 3}\right)^{-{1 \over 3}} \delta^{2 \over 3} \lambda^{-2}\Omega_{b,g} ({M \over \mu m_p})^{1 \over 3} n_0^{2 \over 3}(1 +z)^2,
\end{equation}
yielding
\begin{equation}
N = 1 \times 10^{21} \left({\Omega_{b,g} \over 0.01}\right)^{2 \over 3}\left({M \over 5 \times 10^{11}M_{\odot}}\right)^{1 \over 3}(1 + z)^2\qquad {\rm cm}^{-2}, 
\end{equation}
with $\Omega_{b,g}$ the baryonic mass fraction in the protogalactic
disk relative to the total perturbation. Our numerical estimates are
sensitive to this parameter. In fact, $\Omega_{b,g}$ depends on the merging
history of the dark matter halos and is a function of redshift. Kauffmann
(1996, her Figure 12) presents the redshift distribution of the baryonic
fraction including the conditional probability for halo merging quoted above.
There is a
clear maximum between redshift 2 and 3 for a biasing parameter $1<b<2$. A
value of $b=1-1.5$ seems most consistent with the latest derivations of
Storrie-Lombardi \& MacMahon (1996).
Therefore, we adopt her results in the numerical work presented below.

To characterize the density, we assume that the scale height, $H$, to disk
size, $R$, denoted by $\eta = H/R $, of the collapsed systems is
approximately constant with a
typical value of $\eta \sim 1/100$. It follows that the density in 
the disk is approximately equal to
\begin{equation}
n = \delta \lambda^{-3}\eta^{-1} \Omega_{b,g} n_0 (1 + z)^3
\end{equation}
yielding
\begin{equation}
n = 5 \left({ 100 \over \eta}\right)\left({ \Omega_{b,g} \over 0.01}\right)(1 + z)^3\qquad {\rm cm}^{-3}. 
\end{equation}

The canonical mass value is chosen to be $M = 5 \times
10^{11}M_{\odot}$. Our estimates of the redshift of molecule
formation are only weakly dependent on this choice if we focus on the
bulk of the matter which is contained in the 1$\sigma$ and 2$\sigma$ density
perturbations, $M<10^{12}M_{\odot}$. We also assume
that at high redshift the formation of stars has not been effective in
locking up the baryons. The ratio
$\psi = ({\rm gas}/ {\rm gas} + {\rm stars})$ should be of order unity
for these proto-galactic objects. Of course,
in more evolved systems such as our Galaxy the ratio is only $\sim 10
\%$. The actual frequency distribution of objects of mass $M$ is
determined by the cosmological model in the context of the
Press-Schechter hierarchical clustering picture (c.f.\ White \& Frenk
1991). For a cosmic density parameter $\Omega =1$, it is assumed here
that the total baryonic fraction in galaxies $\Omega_{b,g}$, at a given
redshift, cools to a
disk while the dark matter remains in the halo. The fraction of
baryons in galaxies is typically 10\% of the total number of baryons
available. The cosmic density of baryons $n_{b0}$ is found to be
$n_{b0}= (3 H_0^2/ 8 \pi G )(\Omega_{b} / \mu m_p) = 10^{-6}
(\Omega_{b}/0.1)h^2$, where $H_0$ is the Hubble constant, $ h = H_0
/100km s^{-1} Mpc^{-1}$, $\Omega_b$ is the baryonic fraction of matter
in the Universe and $\mu$ denotes the reduced mass of the primordial
gas for a helium abundance by number of 10\%.

The above formulation is for single objects of a given mass and it is
therefore worthwhile to formulate a statistical criterium which is
more directly related to the initial conditions of structure
formation. Equation (1) can also be written as
\begin{equation}
N=n_0 r_0 \delta^{2/3} \lambda^{-2} \Omega_{b,g} (1+z)^2,
\end{equation}
where $r_0$ is the initial radius and $r_0 = (3 M/4 \pi)^{1/3}$. Using
the Press-Schechter formalism (see White \& Frenk 1991, Equations (1)
and (2)), the abundance of halos $n(M, z)$ with mass between $M$ and
$M + dM$ is given by

\begin{equation}
n(M,z)= ({2 \over \pi})^{1 \over 2} ({\rho_0 \over M^2}) y {d \log y
\over d \log M}\exp ({-{y^2 \over 2}})
\end{equation}
where $y = {\delta_c (1 + z) / \sigma(M)}$, we take $\delta_c=1.68$
and for standard CDM cosmology $\sigma(M)=16.3
b^{-1}(1-0.3909 r_0^{0.1}+0.4814 r_0^{0.2})^{-10}$, where $r_0$ is in
units of Mpc and is given by $r_0 = 1 ( M /10^{12} M_{\odot}) (\Omega
h^2)^{-{2/3}} Mpc$. The function $\sigma(M)$ can be fitted by an
approximate power law $ \sigma = \sigma_0 (M/M_0)^{\alpha}$, leading to
power law distribution functions $n(M, z) \propto M^{-{\alpha - 2}}(1
+ z)$ for the halo mass distribution below a characteristic turnover
mass defined by $ \sigma(M_*(z)) = \delta_c (1 + z)$. Gas cooling and
resultant disk formation can only occur if the cooling time of the
virialized gas in the dark halo at a temperature $T = 2 \times 10^6
(M/10^{12}M_{\odot})(1 + z) K$ is shorter than the dynamical time
at a given redshift. This
leads to an effective cooling mass as a function of redshift denoted
$M_{cool}(z)$. We show in Figure 1 the halo abundance distribution
$n(M,z)$. Also shown in the lower $M-z$ plane of Figure 1 are the
turnover mass $M_*$ and the limiting mass of the dark halo,
$M_{cool}$, in which there is gas, with $\Omega_{b, h} =0.05$, that
can cool at the given redshift (see also Section 5).

\subsection{The Background UV Radiation Field}

The background UV radiation field is well approximated for $0.5<z<2.5$ by
\begin{equation}
J = J_{-21}(1 +z)^{3 + q}
\end{equation}
where $J_{-21}$ is the fiducial value for the intergalactic background
in units of $10^{-{21}}$ erg cm$^{-2}$ s$^{-1}$ sr$^{-1}$ Hz$^{-1}$. The value
of $q$ lies between $0.5 \leq q \leq 1$, and reflects the
uncertainty in the detailed spectral shape of the UV continuum of
quasars. Current observations seem to favor a value $ q = 0.5$ (Tytler
\etal~1995).  For $z<3$, Equation~(5) is accurate to within 30\% over
the 912--2000 \AA\ range which is the relevant wavelength region for
the H$_2$ photodissociation process.  Above a redshift of three, the quasar
population turns over and the metagalactic background decreases with
increasing redshift.

In order to more clearly distinguish the effects of $H_2$ dissociation
and ionization heating, the background radiation field is separated
into two components, one above and one below the Lyman limit at $912 A$,
denoted by $J^{>}$ and $J^{<}$, respectively. The actual calculation of the
radiation field as a function of both wavelength and redshift requires
a numerical integration of the cosmological radiative transfer
equation in the absorption line forest (Haardt \& Madau 1996). We
call this radiation field $J_b$ which is the one used in the numerical
computations.

\subsection{The Stellar Radiation Field}

When star formation is initiated in the protogalaxy, the internally
generated radiation field can become significant. For example, at $1000$
\AA\ the Draine-Habing estimate for the radiation field in our Galaxy,
$J_{D-H}$, gives an effective radiation field of
$J_{-{21},D-H} = 30-50$. We calculate the luminosity, $L_*$, and spectral
energy distribution for
a continuous star formation rate of $1 M_{\odot}yr^{-1}$, with a
Salpeter Initial Mass Function (IMF), and a lower mass cut-off
of $1 M_{\odot}$ (Leitherer \& Heckman 1995).
To obtain an intensity,
we assume that the stars are being formed uniformly troughout a disk with area
$A$ whose nominal value is $100 kpc^2$.
In the numerical results we include the more detailed estimates of
Kauffmann (1996, her Figures 8 and 13) for the redshift-dependent size of
the disk and the radial HI column density distribution.
We denote the stellar radiation field by $J_*$ which is given by
\begin{equation}
J_* =  L_* \left({{\dot S} \over A}\right), 
\end{equation}
where ${\dot S}$ is the actual star formation rate. 
The total radiation field is then
\begin{equation}
J = J_b + J_*.
\end{equation}
Figure 2 presents the shape and magnitude of the overall radiation field 
$J$ for various
epochs and various star formation rates. For ${\dot S}>0.02$ the stellar
component dominates which strongly increases the intensity in the
912-1110\AA\ region where H$_2$ is dissociated. The increase is most 
pronounced for wavelengths longward of the Lyman limit, leading to variations
in the relative contributions to $J^{<}$ and $J^{>}$.

\subsection{Dust and Metallicity Content}

The visual extinction to distant QSOs for damped Lyman Alpha systems
is not more than a few tenths of a magnitude (Fall \& Pei 1994).
Fundamental limits have been obtained on the depletion of elements
like Zn and Cr in damped Lyman Alpha systems at a redshift of order
$\sim 3$. Inferred dust-to-gas ratios, $\xi_{gd}$, are of the order of 
$\xi_{gd}\sim 0.01-0.1 \xi_0$ where $\xi_0$ is the mean galactic value (Pettini
\etal 1995). The metallicities at these high redshifts are down by
approximately the same factor, $Z \sim 1/30 Z_{\odot}$. An accurate
analytical model for the redshift dependence of $\xi_{gd}$ and $Z$ is
difficult to construct due to the essential role played by
star formation. Nevertheless, this dependence should be included since
it is a crucial ingredient in our analysis of molecule formation with
cosmic epoch. In our numerical work, the metallicity is
calculated from the star formation rate in the closed box limit and
also in the limit of the ratio of stars to gas being small i.e. $1 -
\psi = {\rm star}/({\rm star} + {\rm gas}) \sim {\rm star}/{\rm gas} << 1$.
Consequently we can write
\begin{equation}
Z = \left({y {\dot S}t\over M}\right) = \left({2 y {\dot S} \over 3 H_0 M
}\right)(1 + z)^{-{3 \over 2}}, 
\end{equation}
where $y\approx 0.02$ is the yield of metals like C and O
(Woosley \& Weaver 1995). 

The production of dust follows by assuming that a fraction, $ \xi_l$,
of metals is instantly locked up into grains (typically silicate cores). 
Therefore, we find that the dust-to-gas ratio $\xi_{dg}$ is given by
\begin{equation} 
\xi_{dg} = \xi_l Z = \left({\xi_l y {\dot S} t \over M}\right) = \left({2
\xi_l y {\dot S} \over 3 H_0 M }\right)(1 + z)^{-{3 \over 2}},
\end{equation}
As we shall show later, the critical metallicities in our analysis
turn out to be $\sim 0.03-0.1 Z_{\odot}$. Therefore, we have
neglected the significantly lower initial metallicity that may have
come from the formation of the halo.

\subsection{Gas Pressure, Radiation Pressure and Ionization Parameter}

The pressure, $P$, in the mid-plane of the proto-galactic disk can be
estimated by assuming that the mass in the disk is in the gas phase
and using the Archimedean formula $P = \rho g h$ where $g$ is the
gravitational acceleration normal to the disk. For a thin disk $g$ is
given by $g = 2 \pi G \Sigma$ where $G$ is Newton's constant and
$\Sigma$ is the surface density of matter in the disk. Therefore, one
can write
\begin{equation}
P = 2 \pi \mu^2 m_p^2 G N^2, 
\end{equation}
which can be written using equation (1) as
\begin{equation}
P = 2^{-{1 \over 3}}3^{-{2 \over 3}} \pi^{1 \over 3} G \mu^{4 \over 3} m_p^{4 \over 3} \delta^{4 \over 3} \lambda^{-4} \Omega_b M^{2 \over 3} n_0^{4 \over 3} (1 + z)^4,
\end{equation}
with numerical value
\begin{equation}
\tilde {P} = 1200 \left({ \Omega_{b,g} \over 0.01}\right)^2 \left({M
\over 5 \times 10^{11} M_{\odot}}\right)^{ 2 \over 3} (1 + z)^4\qquad {\rm cm}^{-3} {\rm K}.
\end{equation} 
where $\tilde{P} = P /k$, and $k$ is Boltzmann's constant. 
The $(1+z)^4$ dependence of the pressure is slightly misleading due to the
essential role played by $\Omega_{b,g}$. In fact, due to the maximum in the
redshift dependence of the latter, the sizes of protogalactic disks are smaller
at redshifts $z\ge 3$ compared to $z\sim 1$.
Note also that at redshift $z \sim 1$, the pressure calculated from the
weight of the gas is an order of magnitude greater than the thermal
pressure calculated from taking $nT$. In fact, if one takes the radial
dependence of the column density distribution into account and averages
over only the inner 3 kpc of the disk, then the gravitational pressure is
two orders of magnitude larger than the thermal pressure.
This is also the case in our
Galaxy and consequently there must be some turbulent pressure that
holds up the gas (Boulares \& Cox 1990, Norman \& Ferrara 1996). The
turbulent pressure is due to kinetic motions from supernovae and
superbubbles ultimately due to massive star formation. For low star
formation rates the turbulent pressure may not be effective and the
mid-plane pressure acting on the thermal component may increase
leading to a density increase of more than an order of magnitude. This
needs to be kept in mind when discussing the phase diagrams describing
the state of the protogalactic gas.

The pressure, $\tilde{P_h}$, in a virialized halo with a gas fraction
$\Omega_{b, h}$ is given by
\begin{equation}
\tilde{P_h} = 80 \left({ M \over 5 \times 10^{11} M_{\odot}}\right)^{2
\over 3} \left({ \Omega_{b, h} \over 0.05}\right)^{4 \over 3} (1 +
z)^2\qquad {\rm cm}^{-3} {\rm K}.
\end{equation}
We therefore neglect the halo pressure relative to the disk pressure.
More detailed models require the consideration of the pressures and
density profiles in the halo and the role of cooling flows to
precipitate material from halo to disk. These complex issues will not
be addressed further here (c.f.\ Norman and Meiksin 1996) 
 
The radiation pressure is given by
\begin{equation}
P_{rad} = {4 \pi J \nu \over 3 c},
\end{equation}
and we verify that it is always less than the gas pressure in the
proto-disks for each of our calculations.  We define the ionization
parameter used here by $U = J/n$ with $n$ the total number density, and
a numerical value defined by
$U_{-21} = J_{-21}/n$. The dependence of the pressures and the ionization
parameter are shown in Figure 3. The radiation pressure is typically much
less than the gas pressure for the epochs studied here, but is of the same
order of magnitude for $z\approx 0$. The ionization 
parameter varies by a factor of five over the $1<z<3$ range, because of
the turn-over in the quasar population around $z=2.5$.

\section{Gas-Phase Formation of $H_2$ with Background Radiation}

We now turn to examine the epoch when molecular hydrogen is produced
in the gas phase. We first use a simple model (Donahue \& Shull 
1991) to find the approximate analytical expressions for the molecular 
hydrogen abundance through the gas phase $H^{-}$ route.
\begin{equation}
{n(H_2) \over n} = 1.4 \times 10^{-5} T^{0.88} U_{-21} x,
\end{equation}
where $x = n_e/n$ is the electron abundance and is given by
\begin{equation}
x = 0.02 T^{0.42} U_{-21}^{1 \over 2}.
\end{equation}
These equations assume that the H$^{-}$ abundance is in equilibrium and
its dominant destruction channel is the reaction with neutral hydrogen yielding
H$_2$. Conversely, the dominant destruction channels of molecular hydrogen
are UV photodissociation and collisional dissociation at temperatures
$T>2500$ K.
A simple model for the ionization heating yields 
\begin{equation}
T = 3 \times 10^{3} U_{-21}^{0.88} n^{0.12}, 
\end{equation}
and the molecular hydrogen abundance is given by
\begin{equation}
{n(H_2) \over n(H)} = 10^{-2} U_{-21}^{0.55}n^{0.14}.
\end{equation}
For our two component model, where the radiation field is
split into a component longward of 912\AA\ and one shortward of the Lyman
limit, we find the temperature to be
\begin{equation}
T = 3 \times 10^3 (U_{-21}^{<})^{0.27} (U_{-21}^{>})^{0.54} n^{0.12} K, 
\end{equation}
and the molecular hydrogen abundance to be equal to 
\begin{equation}
{n(H_2) \over n(H)} = 10^{-2} (U_{-21}^{<})^{0.85}
(U_{-21}^{>})^{-{0.3}} n^{0.15}.
\end{equation}
In the numerical work, the optically thin H$_2$ dissociation rate assumed in
the above equation has been
corrected for the fact that at abundances of the order of 1
column densities in excess of $10^{19}$ cm$^{-2}$, H$_2$
self-shielding plays a role (see below).

\section{Gas-Phase Formation of $H_2$ with Star Formation}

\subsection{Feedback}

It is generally expected that feedback mechanisms may be important in
galaxy formation (c.f.\ Norman \& Ikeuchi 1996). The following model is
considered here. If the internal star formation generates a dominant
radiation field then the gas will be heated to above $2000 - 3000 K$
and the molecules will be collisionally dissociated. We assume that
the absence of molecular hydrogen formation would strongly limit the
efficiency of star formation. From the simple model used above we
find that for the stellar radiation field the temperature can be
estimated to be
\begin{equation}
T = 7.2 \times 10^3 ({ {\dot S} \over 1 M_{\odot} yr^{-1}})^{0.8} ( {100 kpc^2
\over A})^{0.8} ({ n \over 1 cm^{-3}})^{-{0.88}}, 
\end{equation}
giving a limiting star formation rate of
\begin{equation}
{\dot S_{crit}} = 0.08 ({T \over 1000 K})^{1.25} ({ n \over 1cm^{-3}})^{1.1}({ A \over 100 kpc^2})\qquad M_{\odot} {\rm yr}^{-1}. 
\end{equation}
Using the estimate of the density from equation (2) we find that
\begin{equation}
{\dot S_{crit}} = 0.5 ({T \over 1000 K})^{1.25} \left({ 100 \over
\eta}\right)^{1.1}\left({ \Omega_{b,g} \over 0.01}^{1.1}\right)\left( {
A \over 100 kpc^2}\right)(1 + z)^{3.3}\qquad M_{\odot} {\rm yr}^{-1}.
\end{equation}
For a Schmidt star formation law where ${\dot S} \propto N^p$ with,
$1<p<2$, this corresponds to a critical column density
\begin{equation}
N_{crit} = 1 \times 10^{21} \left( 0.5 ({T \over 1000 K})^{1.25}
\left({ 100 \over \eta}\right)^{1.1}\left({ \Omega_{b,g} \over
0.01}\right)^{1.1}\right)^{1 \over p} (1 + z)^{{3.3 - 2 p
\over p}}\qquad {\rm cm}^{-2}.
\end{equation}
For a protogalaxy of a given mass we find that, using the arguments
outlined in Section 2, the radius of the disk decreases with redshift as
\begin{equation}
R = 10 \left( { M \over 5 \times 10^{11} M_{\odot}}\right)^{ 1 \over 3} \left({ \Omega_{b,g} \over 0.01}\right)^{-{1 \over 3}}(1 + z)^{-1}\qquad {\rm kpc} 
\end{equation}
yielding, with $A = \pi R^2$,
\begin{equation}
{\dot S_{crit}} = 1 ({T \over 1000 K})^{1.25} \left({ 100 \over
\eta}\right)^{1.1} ({ M \over 5 \times 10^{11} M_{\odot}})^{2 \over
3}\left({ \Omega_{b,g} \over 0.01}\right)^{0.43}(1 + z)^{1.3}\qquad
M_{\odot} {\rm yr}^{-1}.
\end{equation}
For star formation to proceed at all, ${\dot S}$ should be a factor of a few
smaller.

\subsection{Bursts of Star Formation}

We have analyzed the feedback effects in terms of a fixed, continuous
star formation rate. We now discuss the burst mode where we envisage
the following cycle: the star formation rate is sufficiently high, so that
all molecules are dissociated; the star formation turns off; the
molecules reform; and another burst occurs.

The time scale, $\tau_D$, for the dissociation of molecular hydrogen,
assuming no self-shielding, is
\begin{equation}
\tau_{D} = 2 \times 10^5 J_{-21}^{-1}\qquad {\rm yr}
\end{equation}
and the timescale, $\tau_F$, for the formation of molecular hydrogen is 
\begin{equation}
\tau_{F} = 4 \times 10^8 \left({1000 K \over T}\right)^{ 1\over 2}\left({ 1 cm^{-3} \over n}\right)\qquad {\rm yr}.
\end{equation}
Consequently, the duty cycle, $D = (\tau_F/\tau_D)$, of on-burst to off-burst is\begin{equation}
D = 8 \times 10^{-3} U_{-21} ({ T \over 1000 K})
\end{equation}
Since the amplitude of the radiation field is set by the level of star
formation rate it follows that in any burst the mass of the stars
formed $M_{*, burst}$ is constant and given by
\begin{equation}
M_{*, burst} = {\dot S} \tau_D \approx 10^6 M_{\odot}.
\end{equation}
Thus, small bursts can occur but with a very low mean effective star
formation rate. It should be emphasized that once a multi-phase medium
has been established, bursts can be sustained (Spaans \& Norman 1996).

\section{The Multi-Phase Structure of the ISM}

\subsection{Numerical Model}

We adopt the heating and cooling curves presented by Donahue \& Shull (1991).
We include cooling by [OI] 63 $\mu$m, [CII] 158 $\mu$m and rotational 
transitions of CO for temperatures below 3000 K. These terms are only of
importance for $z<1.5$. Heating due to photo-electric emission from grains is
included and contributes for $z<1$.
At every redshift Equation (9) is used to determine the metallicity and
expression (8) yields the ambient radiation field. 
The latter is integrated over the hydrogen ionization and H$_2$ 
photo-dissociation cross sections. The ionization, chemical and
thermal balance is solved iteratively until convergence is better than 1\% 
in the temperature and H$_2$ abundance. The redshift dependence of the
baryonic mass fraction is included in the calculation of the column density.
In deriving the H$_2$ abundances we have included the radial N(HI) profiles of
proto-galactic disks as presented by Kauffmann (1996).

\subsection{The Moderate Epoch}

The results for a fixed star formation rate
${\dot S}=0.1M_{\odot}yr^{-1}$ are presented in Figure 4. Between $z=4$ and
$z=2$, the gas is kept at a fairly constant temperature of
approximately 500 K. This confirms the suggestion by Haiman et
al.~(1996) that H$_2$ cooling can facilitate the collapse of high-mass
objects and initiate the formation of structure in the early
universe. It is also apparent that due to the importance of the
stellar radiation field, massive star formation will not occur since a
cool phase of $T<300$ K gas is not present above a redshift of
two. In fact, the existence of this moderate epoch ensures that the
initial cooling time of the disk is not shorter than the dynamical time
at redshifts larger than 3.
If this would not be so, then the star formation rates at high redshifts
would be much too large.

These effects are further reflected in the abundance of CO which remains low
down to a redshift of unity.
When the metallicity reaches a threshold
of approximately $0.03 Z_{\odot}$, the magnitude of the cooling
curve is strongly enhanced. A multi-phase medium is now expected
to result. Note that to simplify Equations (10) and (11) for the metal
enrichment and dust-to-gas ratio development we have assumed a Schmidt
law with index, $p = 1$, implying the star formation rate per unit
mass, ${\dot S} / M $, is a constant.

\subsection{The Phase Transition}

Figure 5 presents phase diagrams for various epochs and star formation
rates. For early epochs, H$_2$ is the only coolant below 3000 K and
only single temperature solutions exist for certain pressures. As the
metallicity increases the $P-n$ curve attains its
characteristic S-shape and lines of constant pressure cut it at three
different temperatures. That is, a multi-phase medium has formed in
which the cool and warm components are stable (Field, Goldsmith \&
Habing 1969) and which can facilitate large scale star formation.
{\it From Figure 5 one finds that the transition to a multi-phase ISM in the
bulk of the galactic disks occurs at $z\approx 1.5$.} At this point it is
timely to mention the recent observations of galaxies in the Hubble
Deep Field presented by Madau et al.~(1996) and Mobasher et al.~(1996) for
the star formation history of the universe.
Although the published results are preliminary,
They seem to suggest evidence for a star formation
epoch around $z\sim 1-2$.

Assuming that there is $1$ supernova per $100 M_{\odot}$ of stars
formed, the filling factor of the hot gas component due to supernovae
is (McKee \& Ostriker 1977)
\begin{equation}
Q = 0.5 E_{51}^{1.28}({ {\dot S} \over 1 M_{\odot} yr^{-1}})({5 \times
10^{11}M_{\odot} \over M}) ({ 0.01 \over \Omega_{b,g}})({n \over 25
cm^{-3}})^{1.14} ({\tilde{P} \over 2.5 \times 10^4 cm^{-3} K})^{-{1.70}}
\end{equation}
which can be written using the critical star formation rate  rate as
\begin{equation}
Q = 0.5 E_{51}^{1.28}({5 \times 10^{11}M_{\odot} \over M})({ 0.01 \over
\Omega_{b,g}})^{0.14}\left({\eta \over 100}\right)^{1.14}({T \over
1000 K})^{-{0.45}}.
\end{equation}
For a Schmidt star formation law, and using Equation (1), we find $ Q
\propto M^{\epsilon} (1 + z)^{-{0.38}}$ where, $0 < \epsilon < 1/6 $,
and consequently the estimate is fairly robust.

Therefore, for high star formation rates at the critical value
(i.e.~more massive objects) the
filling factor of hot gas can be significant in these protogalactic
disks since the pressure is low and the supernova bubbles can expand and fill
large volumes. The above calculation is indicative only since the star
formation is usually clumped and the supernova bubbles generally break out of
the disk and vent their energy into the halo of the protogalaxy. This
process can also result in a feedback that inhibits star formation
(Norman \& Ikeuchi 1989). However high star formation rates a
multi-phase medium may develop where the hot phase is driven by
supernovae energy input. For star formation rates of $0.1-0.3
M_{\odot} yr^{-1}$ and below, the hot phase is not significant and a
two-phase mode can occur.

\section{Formation of $H_2$ on Grains}

An additional important ingredient in the regulation of the star formation 
is the formation
of H$_2$ on dust grains which renders its abundance independent of the electron
fraction. These effects have been included in the models above, but an
analytic model is presented here to facilitate a connection with observations.

In the presence of dust and at low densities, molecular hydrogen is
formed through grain surface reactions. In steady state, the local
H$_2$ density is given by (Tielens \& Hollenbach 1985, van Dishoeck \&
Black 1986)
\begin{equation}
n(H_2)={{A n_H^2}\over{1+2An_H}}, 
\end{equation}
where $n_H=n(H)+2n(H_2)$ is the total hydrogen density. We have
discussed the mean density in the disk but it is necessary to make a
good estimate of the mean density in the diffuse clouds in the cold
neutral medium component where the $H_2$ formation is most likely to
be initiated. We choose to parameterize the clumpiness of the cold
neutral clouds by a diffuse cloud covering factor $ f_c $. If, as we
have indicated above, the protostellar disks are not subject to
substantial massive star formation before molecules are formed then we
do not expect the development of a supernova-driven hot ($\sim 10^6
K$) component. Consequently, the temperature and density contrast
between the cloud and intercloud component will be modest. We infer
that the covering factor of the diffuse cloud component is of order
$\sim 1$.

The controlling parameter, $An_H$, is the ratio of the formation rate of 
molecular hydrogen on grains to the photo-dissociation rate of 
$H_2$.$A$ is given by  
\begin{equation}
 A= {k_sT^{1/2}\xi \over I_{UV}R_{thin}\beta(\tau)
e^{-{\tau_{UV, C}}}}, 
\end{equation}
where $I_{UV}$ denotes the enhancement of the average UV background
with respect to the interstellar field at $z=0$, $\tau_{UV, C}$ is the
optical depth in the UV continuum and $\tau_{UV, C} = 2.5 \xi A_V$,
$R_{thin}$ is the unattenuated $H_2$ photodissociation rate and $\beta$
is the self-shielding function.
The rate constant, $k_s$, depends
on the nature of the grains. The value of $R_{thin}$ depends on the
precise slope of the UV background between 912 and 1100 \AA.

For the present discussion,
we want to determine for which epoch molecular hydrogen self-shields
in the cold diffuse clouds created in the phase transition, and therefore
adopt the constraint
\begin{equation}
A\ge{{2}\over{n_H}}, 
\end{equation} 
yielding a fractional H$_2$
abundance of $0.4$. For self-shielding, the constraint (37) reduces to
\begin{equation}
{I_{UV}R_{thin}\tau_{UV, L}^{-{1 \over 2}} \exp ({- \tau_{UV, C}})} \le
{\left({1 \over 2}\right)k_s T^{1/2} \xi n_H}. 
\end{equation}
Explicitly including the redshift dependence and the dependence on 
covering factor we find that
\begin{equation}
I_{UV, 0}R_{thin} f_c^2 \tau_{UV, L, 0}^{-{1 \over 2}} \exp (- \tau_{UV, C,
0} f_c^{-1} (1 + z)^{1 \over2}) (1 + z)^{{1 \over 2} + q}  \leq
\left({1 \over 2}\right) k_s  T^{1/2} \xi_0 n_{H, 0}.
\end{equation}
Subscripts with zero refer to present time values.

For ${\dot S} \sim 0.03-0.1 M_{\odot} yr^{-1}$ the internally generated
radiation field dominates at lower redshifts, $z \sim 1$.  In the
limiting case where $q =1$, the total extinction varies
slowly with redshift since the lower gas-to-dust ratio of the absorber
compensates for its higher column density. In this case, the critical
redshift for $H_2$ formation, $z_{mol}$, is given by
\begin{equation}
(1 + z_{\rm mol}) \le \left(I_{UV, 0}^{-1}R_{thin}^{-1}f_c^{-2}\tau_{UV,
 L, 0}^{1 \over 2} \exp {(\tau_{UV, C, 0} f_c^{-1})} k_s T^{1/2} \xi_0 
 n_{H, 0}\right)^{1 \over 2}.
\end{equation}
With a numerical value $z_{\rm mol} =1.6$.

Including variations in the value of ${\dot S} \sim 0.1-0.5$, this
yields a fiducial range for $H_2$ formation of, $1.2\lesssim z
\lesssim 2.0$.  These estimates have the advantage that they are
weakly dependent on the input parameters. The strongest dependence is
on covering factor and star formation rate. Obviously, for small
covering factors molecules may form at higher redshift but they will
be less observable. Note that the epoch of significant dust shielding
and cooling are similar since the dust production and metallicity
production are closely related.

The increasing dust columns can absorb a significant
fraction of the internal stellar radiation field and this will influence the
the time scale on which subsequent star formation proceeds.
Using the formalism of McKee (1983) it is possible to estimate the effect of
the dust column on the star formation rate. There is a critical extinction
of approximate 4 mag, necessary to absorb most of the radiation of newly
formed stars. The time scale to convert all available mass into stars 
for a density of roughly 1000 cm$^{-3}$ is given by
\begin{equation}
3.2\times 10^7\quad\xi({{1}\over{2\xi^2}})^{0.5+1.25/A_V}[(1+{{1}\over{8\xi^2}})^{0.5}+({{1}\over{8\xi^2}})^{0.5}]^{1-2.5/A_V}e^{16/A_V}\qquad {\rm yr}
\end{equation}
and the dust extinction $A_V=4\xi(1+z)^2$. For $z>2$ this
time scale is much larger than the local Hubble time, but for epochs later
than 1.7 star formation can proceed efficiently in the bulk of the galaxies.

\subsection{Statistical Approach}

The covering factor discussed above is ultimately determined by the
process of galaxy formation. Using the statistical approach discussed
in Section 2, for any given redshift, the condition (37) for H$_2$
formation on grains can be written as
\begin{equation}
r_0>r_{\rm 0,crit}(z).
\end{equation}
The fraction of galaxies which contain molecules, $f_{\rm mol}(z)$, is 
then given by
\begin{equation}
f_{\rm mol}(z)={\int_{\ell(r_{\rm 0,crit}(z),z)}^{\infty} e^{-x^2} dx 
 \over \int_{\ell(0,z)}^{\infty} e^{-x^2} dx},
\end{equation}
where $\ell(r_0,z)=\left({\delta_c (1+z) \over 2^{1/2} \sigma(r_0)}\right)$.
The definition of the statistical epoch of molecule formation is taken to be
$f_{\rm mol}(z)\approx 0.5$. Inversion of this condition yields 
$<z_{\rm mol}>$. For the parameter values adopted above and putting
the biasing parameter $b=1$, 
the relations $r_{\rm 0,crit}=0.17(1+z)^{2+2\alpha}\quad {\rm Mpc}$ and 
$<z_{\rm mol}>=1.5$ approximately hold.
Note that $r_{\rm 0,crit}$ does not dependent on biasing, but the value of
$<z_{\rm mol}>$ does.

Finally, the probability, ${\cal P}$, that a given line of sight to a
QSO will show molecular absorption by an intervening damped Lyman
Alpha cloud in a (proto-)galaxy is given by (Peebles 1993)
\begin{equation}
{\cal P}\approx\int \sigma n_0 c H_0^{-1} \Omega_b^{-1/2} (1+z)^{1/2}
dz,
\end{equation}
where the integral extends over redshifts between 0 and $<z_{\rm
mol}>$ and $\sigma =\pi r_g^2$. The characteristic radius (bright
part) of the galaxy is chosen equal to $r_g=10h^{-1}$ kpc. With the
parameters given above this yields ${\cal P}\approx 8\times 10^{-2}$.

Once the conditions are satisfied for molecular hydrogen formation the
time scale is of order $ \sim 1/(n_H \xi) Gyr \sim 10^8 (1 +z)^{-(p +
1)}yr $. The ion-chemistry proceeds much faster on a time scale of
approximately $3 \times 10^3 (1 + z)^{-6}yr$. Thus, once the
hydrogen molecules can be formed on grains, the full range of diffuse
cloud species follows immediately.

\section{Conclusions, Discussion and Implications}

Initially motivated by QSO absorption-line observations indicating an
absence of molecular hydrogen at high redshift, we have examined the
formation of molecular hydrogen, after the epoch of re-ionization, in
the context of canonical galaxy formation theory due to hierarchical
clustering. The issue of molecular hydrogen at high redshift has been
discussed in recent interesting papers by Haiman \etal (1996) and
Tegmark \etal (1996). Whereas they concentrated more on the
redshifts above $z \gtrsim 10$, we are focusing on the physics of the
phases of the protogalactic gas at reshifts $z \sim 1-10$. We have
been able to use recently computed accurate models for the cosmic
background radiation field (Haardt \& Madau 1996) and for the
radiation field from a stellar population with a Salpeter IMF
and with a $1 M_{\odot}$ lower-mass cut-off in a continuous star
forming mode (Leitherer \& Heckman 1995).
We have also used a simple model for a
protogalactic disk and followed conventional CDM
cosmology extended by more recent numerical calculations when
considering the spectrum of disk masses and column densities. We have
mainly concentrated on structures associated with 1$\sigma$ and 2$\sigma$
density fluctuations which are expected to form the bulk of the stars.

As with the recent work by Haiman \etal (1996) and Tegmark \etal (1996) we
find there is an initial epoch of $H_2$ production in the gas phase
produced through the $H^-$ channel route where the abundance of
molecular hydrogen is approximately $1\%$ and given simply in terms
of the ionization parameter by equations (20) and (22). Predicting the
details of the state of the interstellar medium of protogalaxies is a
complex task. We have normalized to what is known from work on
the neutral phase of our Galaxy and assume that the fundamental units
in the gas phase of protogalaxies during this epoch are akin to the
diffuse clouds found in our own galaxy. We have shown that until the
metallicity of the gas achieves $Z \sim 0.03-0.1$ at a redshift of 1-2,
cold giant molecular
clouds are not formed due to inefficient cooling. We have found
that star formation in the protogalaxies can become
self-regulated due to heating of the gas by the internal stellar radiation 
field. We have given a simple analytic model for the
feedback process in Section 4.1. It is possible to
define a maximum star formation rate during this epoch. Plausible
estimates give a rate of $ \lesssim 1 M_{\odot} yr^{-1}$ and therefore
we have considered an appropriate fiducial star formation rate to be
$\sim 0.1-0.3 M_{\odot} yr^{-1}$. The production of metals and dust
proceeds slowly in this self-regulated mode. This slow star formation
phase was shown to terminate once the metal abundance increased to a
level of approximately $ Z \sim 0.03-0.1 Z_{\odot}$. From an analysis of
the phase diagrams in Figure 5, we found that species such as C, O and
CO become sufficiently abundant and can cool the gas below $300 K$ to
$\sim 10 -30 K$. At this point a phase transition can occur in the
protogalactic gas. For the low fiducial star formation rates discussed
above, we find this to be a transition to a two-phase medium as
described by Field, Goldsmith \& Habing (1969). Dense molecular clouds can
form and the star formation is no longer self-regulated in the
manner described above since the UV radiation does not penetrate the
dense cores of the clouds. We expect that rapid, massive star
formation ensues and the abundance of metals and dust increase
concomitantly. The dust abundance also becomes sufficiently high to
allow molecular hydrogen formation on grain surfaces. With the
increased star formation rates, the ISM will change to one dominated by
supernovae energy input (McKee \& Ostriker 1977) with significant
exchange of mass, energy and metallicity from the disk to the halo
(Norman \& Ikeuchi 1989). In a subsequent paper we will investigate in more
detail the effects of this phase transition on the evolution of dwarf
galaxies and the importance of metal loss driven by supernova explosions.

Our analysis may be of relevance to the G-dwarf problem: the Simple model
(Pagel 1989, and references therein) for chemical enrichment overproduces
the number of metal-poor stars
(Cowley 1995; Worthey, Dorman \& Jones 1996 and references therein). Of the
many solutions proposed
for the G-dwarf problem, the simplest appears to be that by the time a few
percent of the gas mass of a galaxy is assembled into stars, the remaining
gas reservoir is
already enriched in metals and has not yet experienced any star formation.
The moderate phase due to feedback identified in this work,
is likely to cause star formation in an inhomogeneous way. The very nature
of the feedback mechanism dictates that star formation in one location
strongly suppresses additional star formation in its vicinity.
The subsequent phase transition
then causes rapid star formation throughout the
mostly unprocessed ISM which now has a metallicity close to 10\% of solar.

In our analysis we have assumed a fixed value for the collapse factor
$\lambda^{-1}$. In reality, the distribution of spin parameters
may be quite wide (Warren et al.~1992; Dubinsky \& Carlberg 1991).
If we view the disk as being formed from a spherical object sustaining a
low H$_2$ abundance, driven by the extragalactic UV background, then an
increase in the collapse factor will increase the total and H$_2$ column
densities quadratically and the local
density like $\lambda^3$. Consequently, more of the stellar radiation can be
absorbed locally, preserving the H$_2$ abundance and suppressing the
global heating, and a higher star formation rate can be sustained. We
estimate, although tentatively, that for objects with
$\lambda\sim 0.02$, a factor of 3.5 below our canonical value, the redshift
at which the phase transition occurs, can be as high as three.
More detailed knowledge of the $\lambda$ distribution is necessary to assess
how common such high redshift objects are.

In summary, from our elementary cosmological model we conclude that this
new mode of star formation, where objects now akin to giant molecular
clouds in our Galaxy become the sites of star formation, occurs at a redshift
of approximately 1.5 with a value higher by a factor of 2 if more massive
initial perturbations or larger collapse factors are considered. The phase
transition in the interstellar medium of
protogalactic disks as analyzed in this paper is now a plausible
physical reason that the formation of disks of galaxies occurs at a
redshift of order unity with a significant increase in star formation
after the metallicity has achieved a value of order $Z \sim 0.03-0.1
Z_{\odot}$. These findings are consistent with the recent studies of
the Hubble
Deep Field (cf.~Madau et al.~1996; Mobasher et al.~1996). The combination
of feedback and a phase transition can provide a natural solution
to the G-dwarf problem.

We are grateful to Tim Heckman, David Neufeld and Rosemary
Wyse for illuminating and stimulating discussions that contributed
significantly to our understanding, and to Andrea Ferrara, Claus
Leitherer and Piero Madau for such discussions and also for providing
us with their excellent data on low-metallicity cooling curves, the
spectrum of radiation from stellar populations and the cosmic
background radiation field. We are also grateful to Piero Rosati for his
assistance in the presentation of the numerical results.
We thank the anonymous referee for his detailed and valuable comments.
MS also acknowledges with gratitude the support of NASA grant NAGW-3147
from the Long Term Space Astrophysics Research Program.


\clearpage

\begin{figure}
\caption{Figure 1 shows the abundance of halos as a function of mass
and redshift for standard CDM with biasing parameter $b=1$.  The lower
$M-z$ plane shows, as functions of redshift, the characteristic
turnover mass, $M_*$ (solid curve), in CDM and the limiting mass of
the dark halo, $M_{cool}$ (dashed curve), in which gas can cool. }
\end{figure}

\begin{figure}
\caption{Figure 2 shows the combined stellar and 
background radiation fields
at redshift $z = 1$, $2.5$, and $4$ for star formation rates of $0.01$, 
$0.1$ and $1 M_{\odot} yr^{-1}$. For rates larger than 0.03 the stellar 
component dominates and feedback is important.   }
\end{figure}

\begin{figure}
\caption{Figure 3 shows the derived gas pressure,
radiation pressure and ionization parameter as functions of
redshift for ${\dot S}=0.1 M_{\odot} yr^{-1}$. 
The radiation pressure is typically much less than the gas
pressure due to the low star formation rate at which the feedback
mechanism is most efficient.   }
\end{figure}

\begin{figure}
\caption{Figure 4 shows the results of the
calculations for the gas temperature and the abundance of molecular
hydrogen, CO, dust ($\xi_{gd}$), and metals ($Z_m$) as functions of redshift 
for a star formation rate ${\dot S}=0.1 M_{\odot} yr^{-1}$.   }
\end{figure}

\begin{figure}
\caption{Figure 5 shows pressure, $P$, and gas density, $n_H$,
phase diagrams for redshifts 1, 2 and 3, respectively and star
formation rates of $0.01$, $0.1$ and $1 M_{\odot} yr^{-1}$. The curve
in the upper left panel denotes constant pressure. The symbol $Z$ indicates
the metallicity with respect to galactic at that epoch for a particular 
star formation rate. The labels F, G and H denote the
thermal equilibria of which G is unstable.   }
\end{figure}

\setcounter{figure}{0}

\newpage
\begin{figure}
\centerline{\psfig{figure=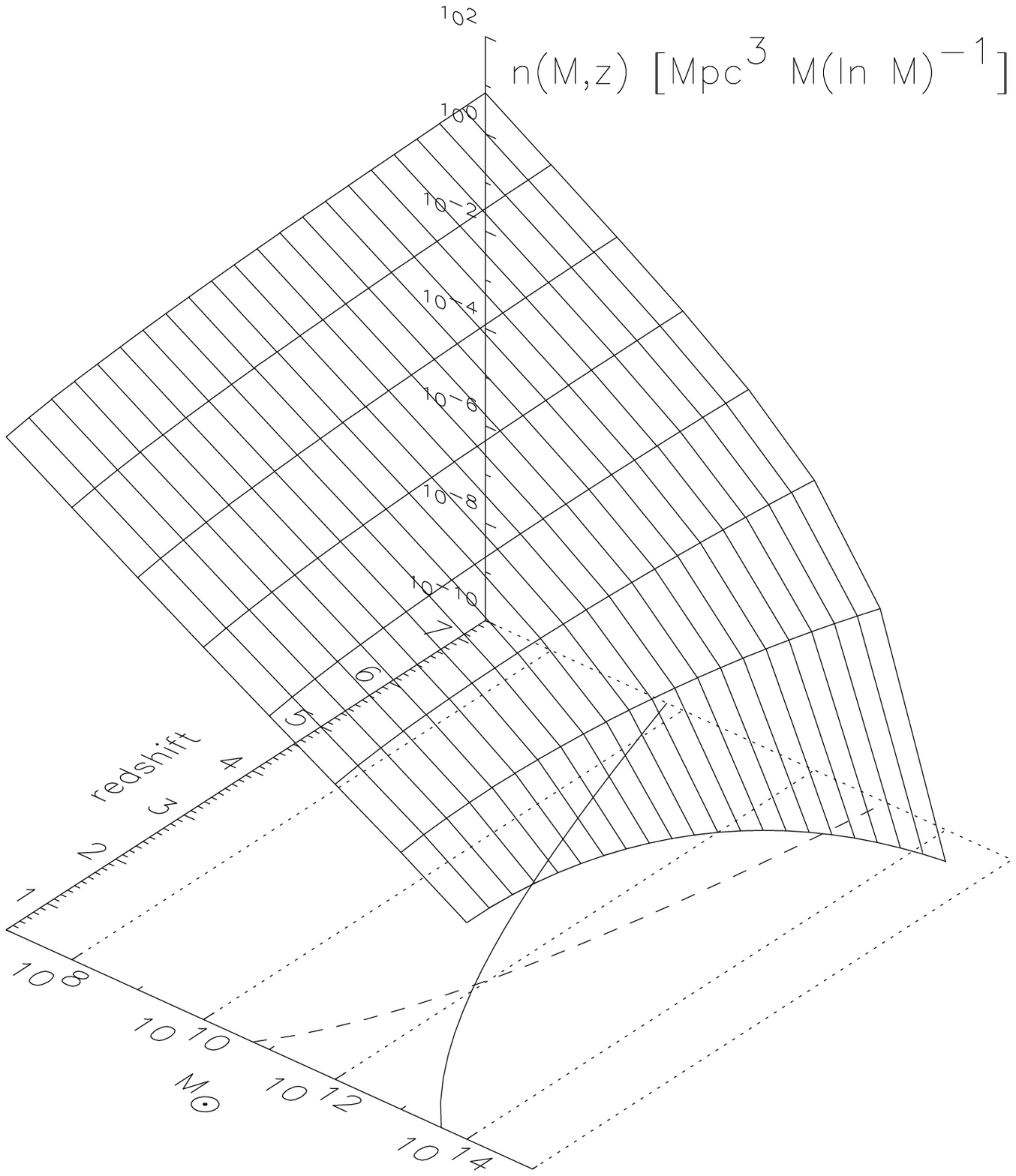,width=6in,angle=-0}}
\caption{\label{figure1}}
\end{figure}

\newpage

\begin{figure}
\centerline{\psfig{figure=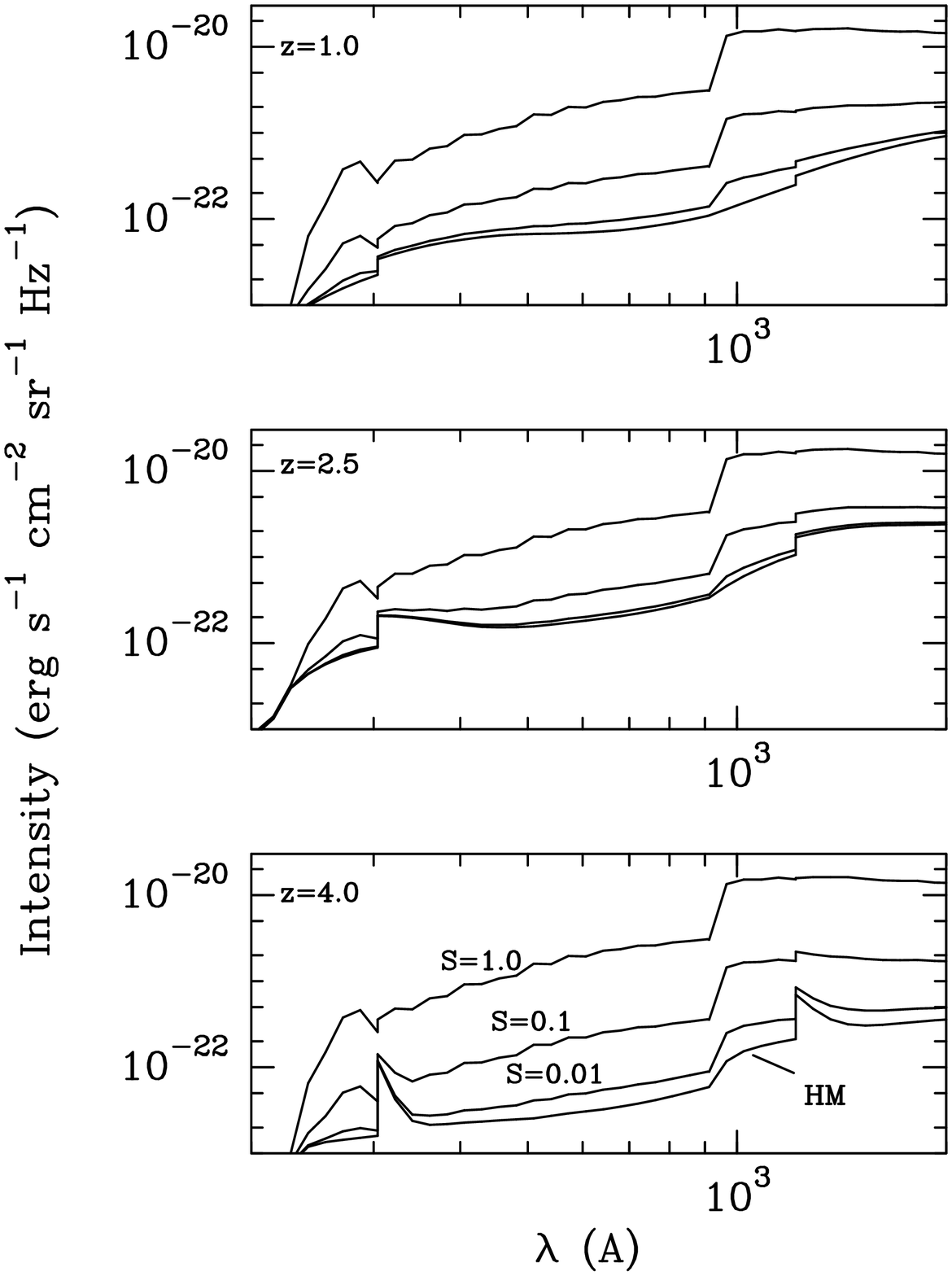,width=6in,angle=-0}}
\caption{\label{figure2}}
\end{figure}

\newpage

\begin{figure}
\centerline{\psfig{figure=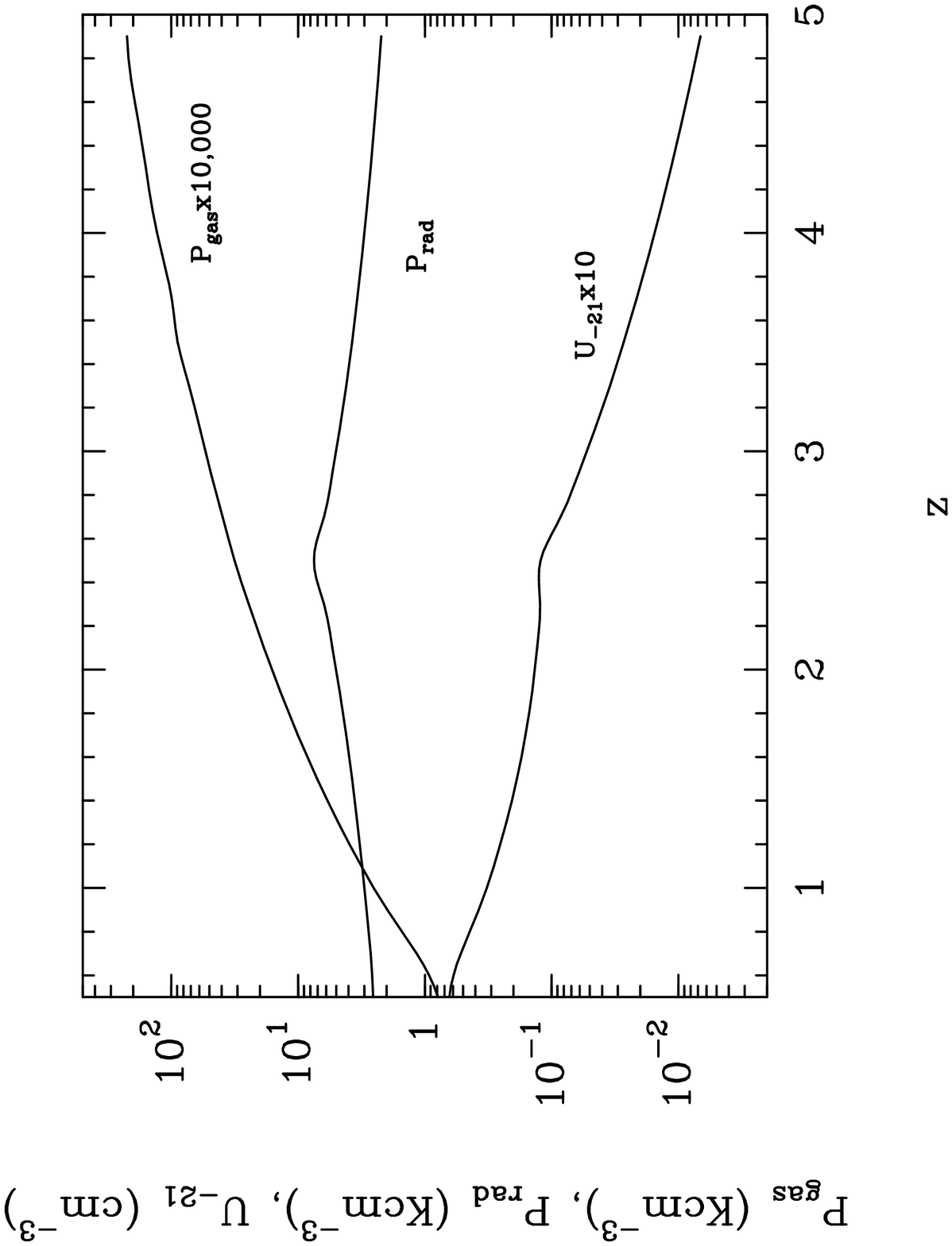,width=6in,angle=-0}}
\caption{\label{figure3}}
\end{figure}

\newpage
\begin{figure}
\centerline{\psfig{figure=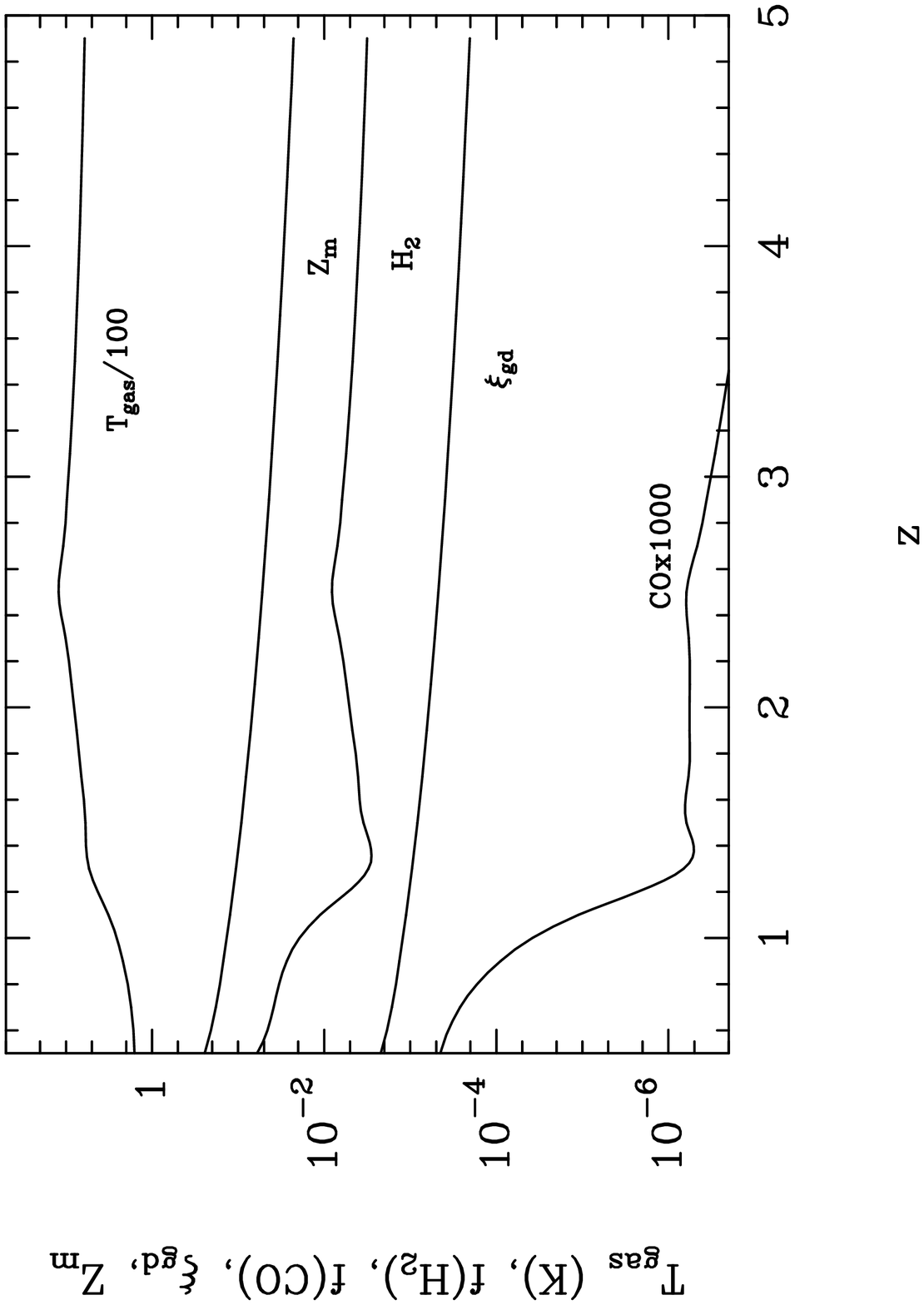,width=6in,angle=-0}}
\caption{\label{figure 4}}                     
\end{figure}

\newpage

\begin{figure}
\centerline{\psfig{figure=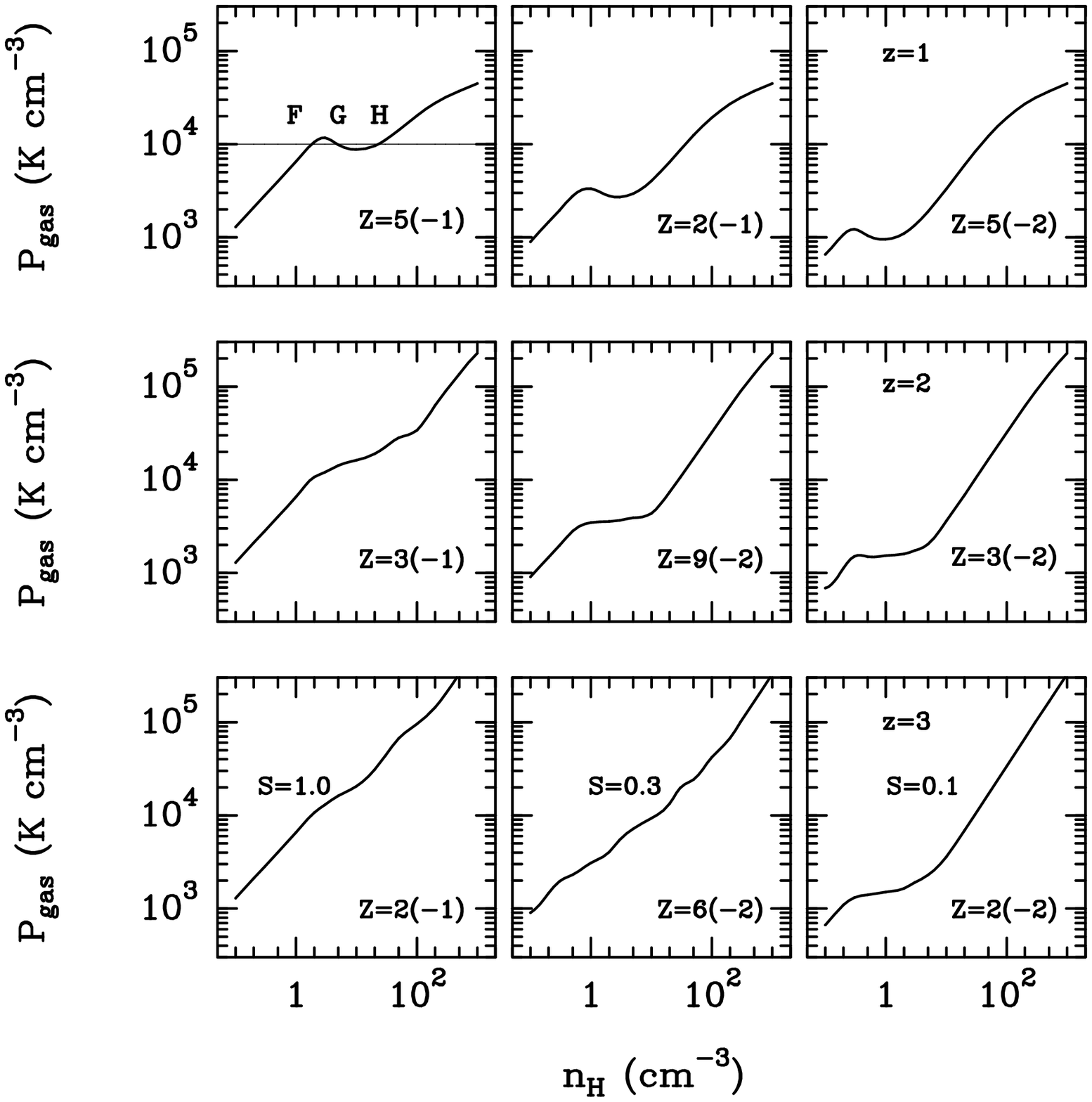,width=6in,angle=-0}}
\caption{\label{figure 5}}
\end{figure}

\end{document}